\documentstyle[12pt]{article}
\input psfig
\def\epm{$e^+e^-$}
\def\vub{$\left| V_{ub}\right|$}
\def\dec{\rightarrow}
\def\vcb {$\left|V_{cb}\right|$}
\def\ups{$\Upsilon (4{\rm S})$}
\def\qsq{$q^2$}
\def\vubc{$\left|V_{ub}/V_{cb}\right|$}

\def\etal{et al.}

\def\bbar{$B^0\bar{B}^0$}
\def\kkbar{$K^0\bar{K}^0$}
\def\etal{{\it et al.}}
\begin{document}
\begin{titlepage}
\rightline{\vbox{\hbox{HEPSY 96-2}
\hbox{October, 1996}}}

\vskip 4.0cm

\centerline {\bf $B$ Decay Studies at CLEO \footnote{Talk presented at the 4th International Workshop 
on B Physics at Hadron
Machines} }
\normalsize
 
\vskip 2.0cm
\centerline{Marina Artuso}
\centerline{Department of Physics,}
\centerline{Syracuse University,}
\centerline{Syracuse, New York 13244--1130}
\centerline{\it e-mail: artuso@physics.syr.edu}
\vskip 1.0cm
\begin{abstract}
Weak decays of heavy flavored hadrons are sensitive probes of several 
facets of the Standard Model. In particular the experimental study of  
$B$ meson semileptonic decays
is starting to pin down the quark mixing parameters in the Cabibbo Kobayashi
Maskawa matrix. 
In addition, some features of the non--perturbative regime of the
strong  interaction are 
probed by these decays. 
New results from the CLEO experiment at the CESR \epm\ collider, based 
on a data 
sample of up to 3.5 fb$^{-1}$  provide crucial 
information on both of these aspects of heavy flavor phenomenology.
\end{abstract}
\vfill
\end{titlepage}
\newpage
\section{Introduction}

Weak decays of heavy flavored hadrons are an excellent laboratory to study 
the Standard Model. In particular B meson decays provide a wealth of 
information 
on the quark mixing elements. While the next generation of high luminosity 
facilities 
have a good chance to measure $CP$ asymmetries in these decays and 
perhaps shed
some light on the puzzling and fundamental phenomenon of $CP$ violation, 
good 
progress in measuring some parameters describing quark mixing has been 
achieved.

In the framework of the Standard Model the gauge bosons, $W^{\pm}$, 
$\gamma$ and 
$Z^o$ couple to  
mixtures of the physical $d,~ s$ and $b$ states. This mixing is described
by the Cabibbo-Kobayashi-Maskawa (CKM) matrix:
\begin{equation}
V_{CKM} =\left(\begin{array}{ccc} 
V_{ud} &  V_{us} & V_{ub} \\
V_{cd} &  V_{cs} & V_{cb} \\
V_{td} &  V_{ts} & V_{tb}  \end{array}\right).
\end{equation}
A commonly used approximate parameterization was originally proposed by 
Wolfenstein \cite{wolf}. 
It reflects the hierarchy between the magnitude of matrix elements 
belonging to different 
diagonals. The 3 diagonal elements and the 2 elements just above the 
diagonal
are real and positive. It is defined as:
\begin{equation}
V_{CKM} =\left(\begin{array}{ccc} 
1-\lambda ^2/2&  \lambda &  A\lambda ^3(\rho -i\eta)\\
-\lambda & 1-\lambda ^2/2 & A\lambda ^2\\
A\lambda ^3(1- \rho -i\eta)&  -A\lambda ^2& 
1\end{array}\right).\end{equation}

$B$ decays probe several of these matrix elements. The study of 
semileptonic decays
allows the measurement of  $\left| V_{cb}\right|$ and 
$\left| V_{ub}\right|$.  In addition, the ratio
${\cal B}(B\dec \rho (\omega)\gamma )/ {\cal B}(B\dec K^{\star} \gamma )$ is
considered a promising avenue to  measure the
ratio $V_{td}/V_{ts}$ \cite{ali}. 

The Standard Model parameterization of the quark mixing via the CKM 
matrix element 
accomodates a complex phase, and therefore offers a 
natural way to model the intriguing phenomenon of $CP$ violation. So far this 
violation has 
been measured only in neutral $K$ decays, and yet it may 
very well be
at the origin of the matter dominated universe that exists now. 

\begin{figure}[htbp]
\vspace{-.8cm}
\centerline{\psfig{figure=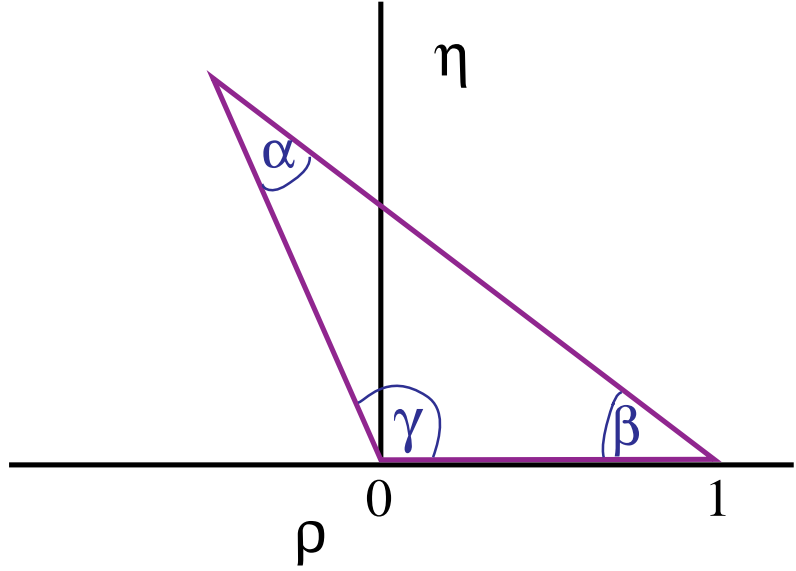,height=4in,bbllx=0bp,bblly=450bp,bburx=600bp,bbury=750bp,clip=}}
\vspace{-.6cm}
\caption{\label{ckm_tri}The CKM triangle shown in the $\rho-\eta$ plane. The
left side is determined by $|V_{ub}/V_{cb}|$ and the right side can be
determined using mixing in the neutral $B$ system. The angles can be found
by making measurements of CP violation in $B$ decays.}
\end{figure}

The CKM matrix must be unitary and the relation between elements of different
rows dictated by this property can be graphically represented as
so called `unitarity triangles'.   Fig.~\ref{ckm_tri} shows the most promising 
of the triangles: the angles $\alpha$, $\beta$ and 
$\gamma$ are all 
related to the single phase in the $CKM$ matrix element.
The study of $B$
decays will eventually
allow the 
measurements of  all the three angles. Thus, it will pose a 
serious challenge 
to the Standard Model description of CP violation and perhaps shed 
some light on 
phenomenology beyond the Standard Model. 

The statistical accuracy corresponding to the present data sample 
accumulated with the CLEO detector, about 3.5 ${\rm fb}^{-1}$ for the 
results 
presented
in this paper, is not yet sufficient to probe $CP$ violation in $B$ decays or 
to
measure rare decays like $B\dec \rho \gamma$  accurately, but is adequate 
to give 
extremely valuable information on the matrix elements $V_{cb}$ and 
$V_{ub}$. 
These data allow us to put  better constraints on the fundamental 
parameters in two 
major ways. On one hand, the measurements reported in this paper 
provide new 
information that reduces the experimental errors on the parameters. 
On the other 
hand, the experimental data provide 
constraints to the theoretical models that are needed to relate measured 
observables to
the fundamental parameters of the Standard Model.

In addition, heavy flavored meson decays are a laboratory to probe the 
strong interaction 
in different dynamical domains. $B$ meson decays probe a 
regime not fully 
amenable to  perturbative QCD calculations, but suitable for the application 
of effective 
theories derived from QCD in some asymptotic conditions. In particular, an approach 
that has 
generated  a lot of interest in the last few years is the so called `Heavy 
Quark Effective 
Theory' (HQET) \cite{hqet}, where the asymptotic behavior corresponds to taking the 
limit 
$m_Q\dec\infty$. New 
data  shedding some light on the hadronic matrix element will be discussed.
 
\section{The Cabibbo Kobayashi Maskawa Matrix and $B$ meson 
semileptonic decays}
The CKM parameters \vcb\ and \vub\ have been studied extensively by the 
CLEO 
collaboration through the study of semileptonic decays $B\dec X \ell 
\bar{\nu}$. The experimental study of semileptonic 
decays has 
addressed both inclusive measurements, where the recoiling hadronic state 
is not 
identified, and exclusive measurements, where the recoiling hadron is 
reconstructed 
through one of its decay channels. 

Inclusive decays have provided several interesting results. Most notably the 
study of the 
end--point of the lepton spectrum, where no charmed hadron final states are 
kinematically 
allowed, has provided the first conclusive evidence of a non zero value of  
\vub\ \cite{bucleo}. 

Exclusive decays provide complementary information. In particular the 
semileptonic 
channel more extensively studied so far is $B\dec D^{\star} \ell \bar{\nu}$. This 
channel is attractive from an experimental point of view because the slow 
$\pi$ emitted in 
the decay $D^{\star} \dec D\pi$ provides a unique signature of this 
hadronic final state. 
In addition, a sharpened attention to this decay has been prompted by the 
suggestion \cite{neubone} from HQET that its study
provides a `model independent' method to determine \vcb .
The arguments for this claim will be discussed below. 

\subsection{The CKM element \vcb\ and the decay $B\dec D  \ell \bar{\nu}$}
The decay $B\dec D  \ell \bar{\nu}$ is interesting for 
several reasons. Analyses similar those for the decay $B\dec D^{\star}  \ell 
\bar{\nu}$ 
 provide ways to understand the
systematic uncertainty in the extraction of the parameter \vcb .

 The hadronic matrix element involved in this decay, the main source 
of uncertainty 
in extracting \vcb\ from the experimental data, is probed effectively by the 
differential decay 
width $d\Gamma /dq^2$:
 
\begin{equation}
\frac{d\Gamma }{dq^2}=\frac{G_F^2\mid V_{cb}\mid ^2 K^3 M_B^2}{24\pi 
^2}\mid f_+ (q^2) \mid ^2.
\end{equation}
where $M_B$ and $M_D$ are the $B$ and $D$ meson momenta 
respectively, $q^2$ is 
the invariant mass of the lepton-neutrino pair and $K$ is the $D$ 
momentum in the $B$ 
rest frame and is given by:
\begin{equation}
K = \frac{M_B}{2}\left\{ \left[ 1 -\frac{M_D^2-q^2}{M_B^2}\right] -
4\frac{M_D^2q^2}{M_B^4}\right\}^{1/2}.
\end{equation}

$f_+(q^2)$ is the form factor describing the hadronic interaction and must 
be extracted 
from theory. Most quark model calculations assume a $q^2$ dependence 
for the form 
factor and predict the normalization at a given kinematic point. The 
normalization factor is calculated either at $q^2=0$ or $q^2=q^2_{max}$. In 
general the 
arbitrariness 
of the assumed $q^2$ dependence is a reason for some concern, but in this 
specific decay 
the $q^2$ range is not very big and therefore we would not expect a strong 
model 
dependence. 

CLEO has recently studied the decay $B\dec D \ell \bar{\nu}$  with 
two different 
techniques \cite{dlnuvarsaw}. In the first method only the lepton and the 
$D$ candidates 
in the final state are found, using the decay $D^+\dec K^-\pi^+\pi^+$. 
Because the 
$B\bar{B}$ pairs are produced nearly  at  rest, the missing mass squared 
$MM^2$ is 
calculated as:
\begin{equation}
MM^2= E_{\nu }^2-\vec{p}_{\nu}~^2\approx
(E_B-E_D-E_{l})^2- (\vec{p}_{l}+\vec{p}_{D})^2.
\end{equation}
Here the approximation consists of assuming $\vec{p}_B\approx  0$, 
relying upon the 
low  magnitude of the $B$ momentum because of the vicinity of the \ups\  
to the threshold  
for $B\bar{B}$ meson production.  The second approach exploits the  
hermeticity of 
the CLEO II detector  and infers the $\bar{\nu}$ momentum from a full 
reconstruction of 
the semileptonic decay.  Stringent cuts need to be applied in this case to 
insure that 
no other sources of missing 4-momentum, like additional $\nu$'s or 
$K_L$'s, are present  
in the event. 

\begin{figure}[htbp]
\vspace{-4.cm}
\centerline{\psfig{figure=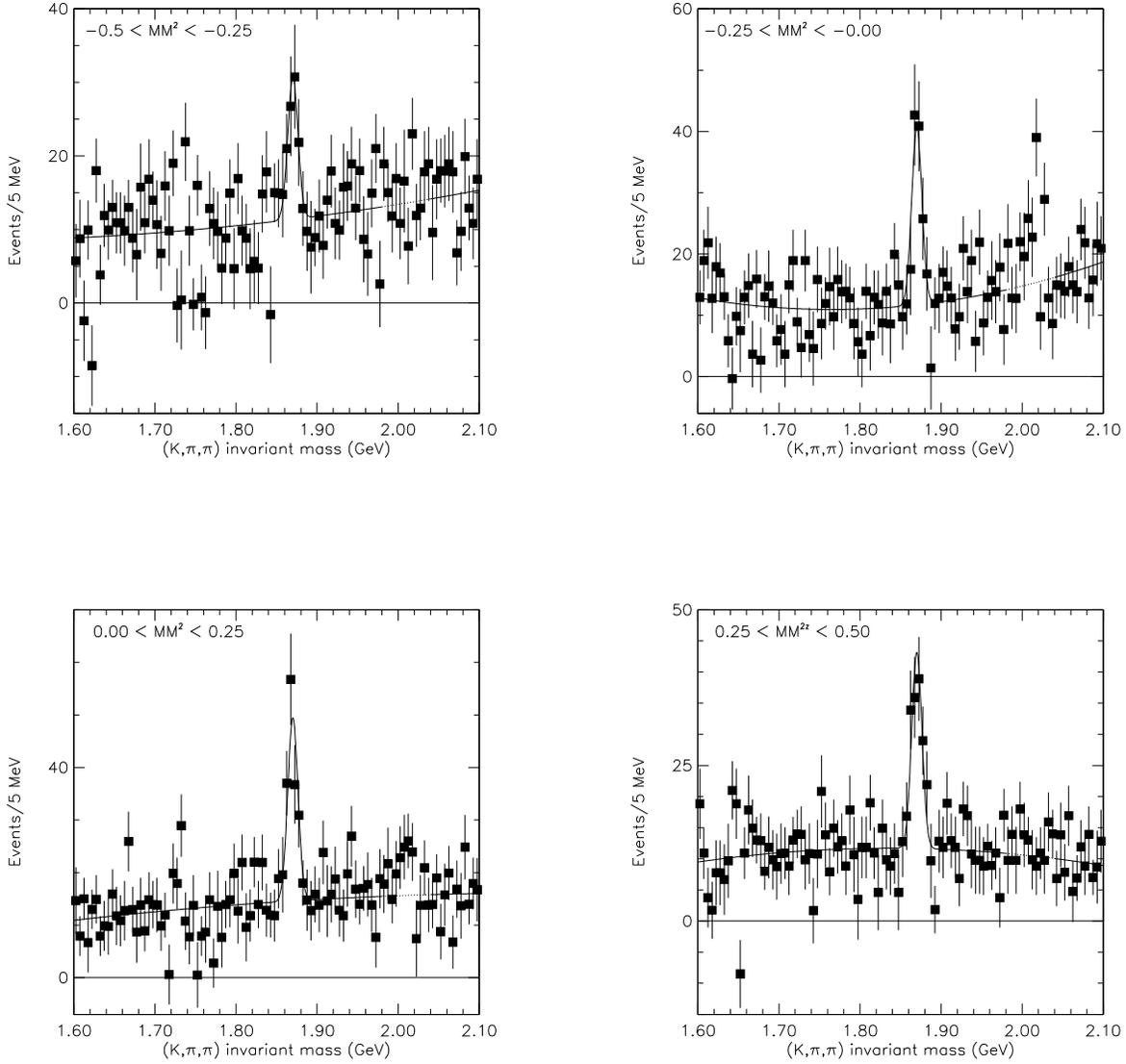,height=7in,bbllx=0bp,bblly=200bp,bburx=700bp,bbury=700bp,clip=}}
\vspace{-.8cm}
\caption{\label{mkpipi}Invariant $K^-\pi^+\pi^+$ mass spectra from CLEO for 
events with an opposite sign lepton in the interval $4>q^2>2$ and for
different $MM^2$ slices.}
\end{figure}

In the former analysis, the $MM^2$ distribution of candidate events 
containing a  $D^{\pm}$ and an opposite sign lepton is studied in
6 different $q^2$ bins GeV$^2/c^2$,  evenly spaced between  0 and 12  
GeV$^2/c^2$. Fig.~\ref{mkpipi} shows the $K^- \pi  ^+ \pi^+$ invariant mass for 
the interval $2< q^2< 4\ {\rm GeV^2/c^2}$.

\begin{figure}[htbp]
\vspace{-.5cm}
\centerline{\psfig{figure=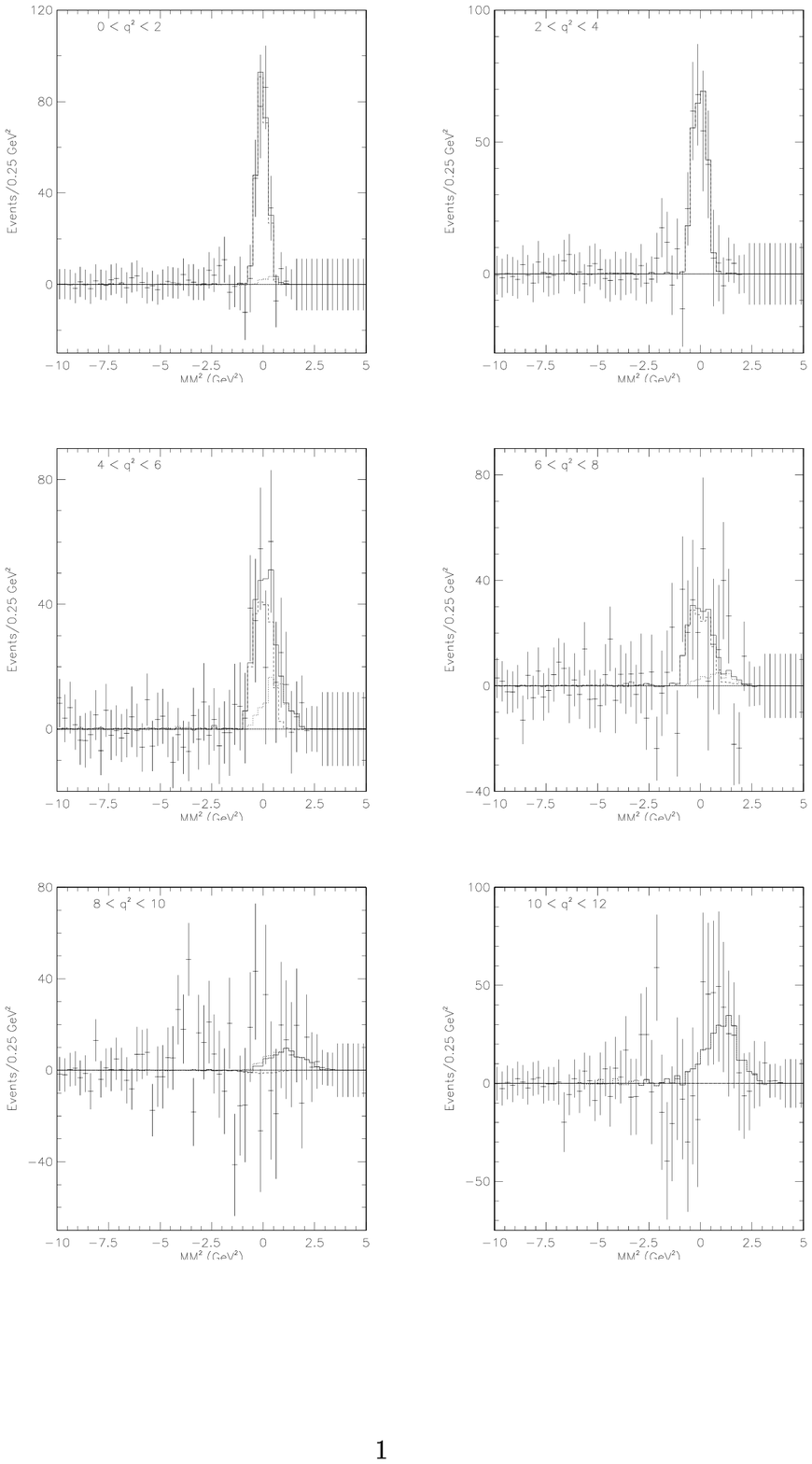,height=7.in,bbllx=0bp,bblly=100bp,bburx=600bp,bbury=700bp,clip=}}
\vspace{.3cm}
\caption{\label{allqsq}The $q^2$ distribution for
 $\bar{B}^0 \dec D^+ \ell \bar{\nu }$ from
the $MM^2$ analysis.\hfill}
\end{figure}

One of the crucial elements of this analysis is an accurate background
estimate. Several sources are subtracted directly, 
using independent control samples. In particular, a major 
 background is the decay $B\dec D^{\star +} X \ell \bar{\nu}$, where 
the  $ D^{\star +}$ decays into the final state $D^+\pi ^0$.  This 
component is subtracted by measuring the $MM^2$ distribution for 
identified $B\dec D^{\star +} X \ell \bar{\nu}$ decays, rescaled by 
the ratio in detection and reconstruction efficiency for the two channels. 
The data sample remaining after direct background subtraction has two 
components: the signal final state $\bar{B}^0\dec D^+  \ell \bar{\nu}$ and final 
states of the kind $B\dec D^+ X  \ell\bar{\nu}$, where $X$ is a 
hadronic final state not coming from the $D^{\star }$ mode discussed 
above. This last background is subtracted by fitting its contribution with the 
shape given by a Monte Carlo simulation. The results of the fit in several 
$q^2$ bins is shown in Fig.~\ref{allqsq}. Note that in different $q^2$ bins 
the 
relative weight of signal and background is quite different. Therefore the 
study of the $MM^2$ in different $q^2$ regions is quite effective in 
isolating the signal from this last background contribution.

\begin{figure}[htbp]
\vspace{-2.1cm}
\centerline{\psfig{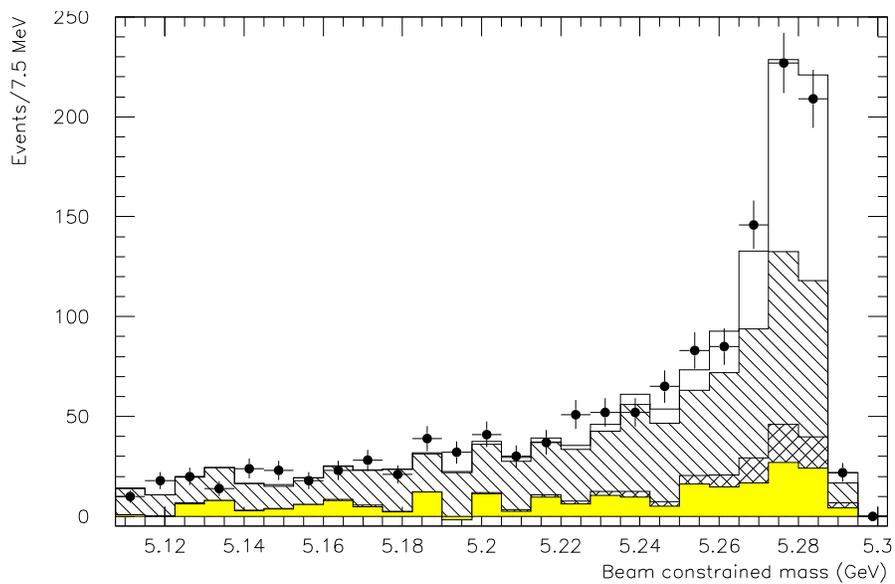}}
\vspace{-.05cm}
\caption[]{\label{dplnunu} Beam constrained mass spectrum for 
$\bar{B}^0 \dec D^+ \ell \bar{\nu}$ with the $\bar{\nu }$ reconstruction
analysis. The white area represents the signal events, the hatched ares
represents the combinatoric background, the crosshatched area represents the
$D^{*+}\ell^-\bar{\nu}$ and the shaded area represents all the remaining
backgrounds.}
\end{figure}

The second technique reconstructs the $\nu$ four--vector by 
summing all the charged track and photon momenta in the event. Since 
the total energy in the event is equal to the center of mass energy and the 
total momentum is zero, the $\nu$ is assumed to have energy equal to 
the difference between center of mass energy and total reconstructed 
energy and momentum equal and opposite to the reconstructed momentum. 
In order to achieve adequate resolution, stringent event selection criteria are 
applied to avoid smearing due to additional undetected neutral particles or 
particles outside the detector acceptance. Consistency between 
the reconstructed energy and momentum  is required. Once $\vec{p}_{\nu 
}$ is estimated, the $B$ meson candidate can be reconstructed with the 
usual procedure for exclusive hadronic final states, as shown in
Fig.~\ref{dplnunu}.

The $MM^2$ technique gives a branching fraction of $(1.75\pm 0.25 \pm 
0.20)$\%, the $\nu$ reconstruction technique  gives $(1.89\pm 0.22 \pm 
0.35)$\% with a combined (preliminary) branching fraction of  $(1.78\pm 
0.20 \pm 0.24)$\%. The statistical errors in these two techniques are 
essentially uncorrelated, while the systematic error is strongly correlated.

The $q^2$ distribution for the $MM^2$ method is shown in Fig.~\ref{qsq_dp}. 
The 
intercept at $q^2 = 0$ is proportional to $\left| V_{cb}f_+(0)\right| $. The 
curve is fitted to the functional form:
\begin{equation}
f_+(q^2) = \frac{f_+(0)}{1-q^2/M_V^2},
\end{equation}
where $f_+(0)$ is the normalization at $q^2=0$, and $M_V$ is the mass of the 
pole. 
The quark model calculations  predict 
$f_+(0)$, and the pole corresponds
the vector meson exchanged in the t-channel, in this case the $B^{\star }$.
The data are fitted including both $f_+(0)$ and $M_V$ as fit parameters, except
in the case of the model developed by Demchuck {\it et al.} \cite{demc}, where
 the mass of
the pole is a definite prediction of the theory.
The results and a comparison with different 
models \cite{ks}, \cite{wbs}, \cite{demc} are given in Table 1. 
Even if this study considers only a restricted set of models, 
the limited range of $q^2$ spanned by this decay makes the results quite
general.  

The first errors in the average value of \vcb\ is the quadrature of the 
statistical and systematic errors in the data and the fact that the fraction of 
neutral $B^0\bar{B}^0$ final state produced at the \ups\ is known only as 
$0.49\pm 0.05$ \cite{stone}. The second error is due only to model 
dependence. 

\begin{figure}[htbp]
\centerline{\psfig{figure=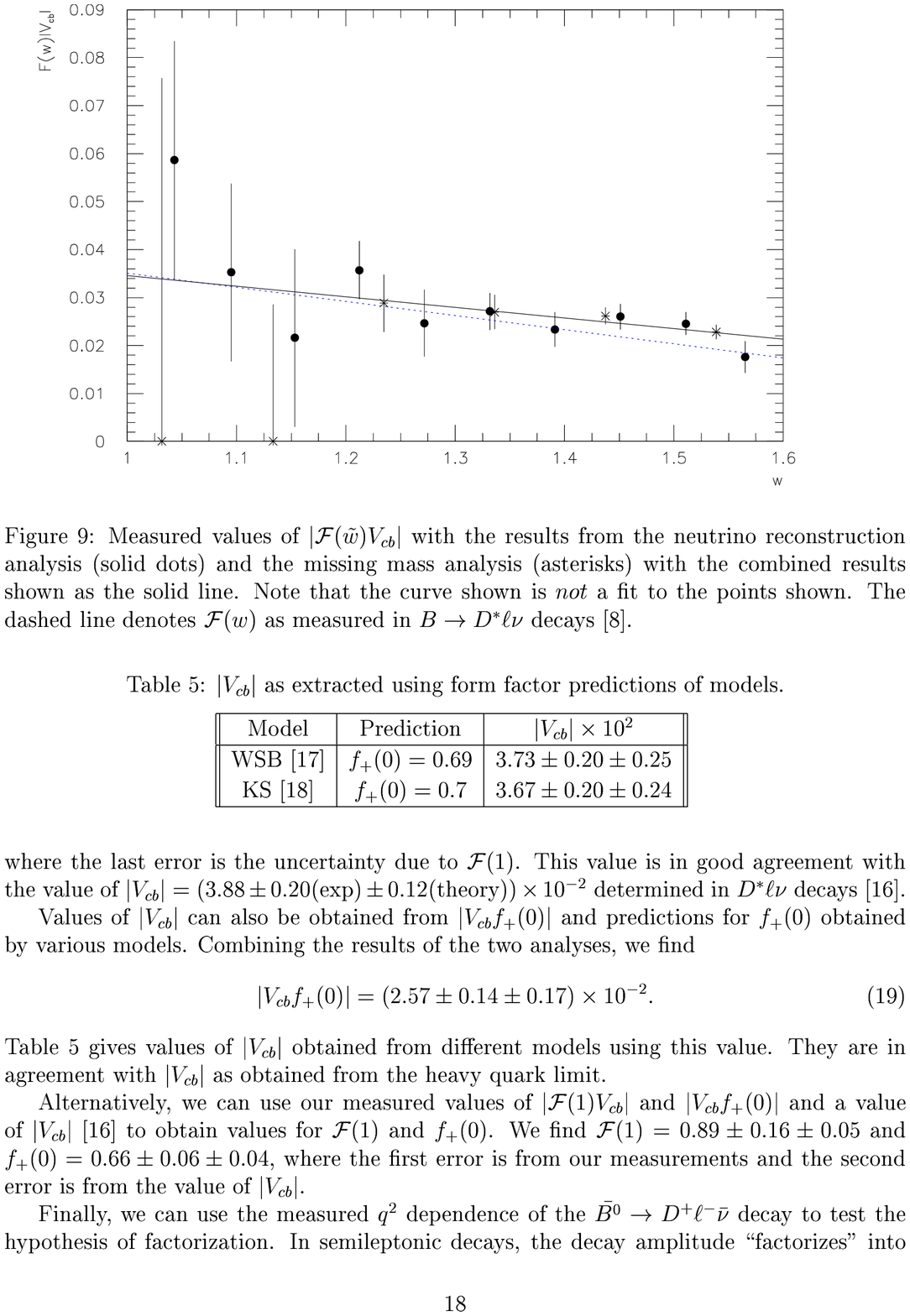,height=5.in,bbllx=0bp,bblly=460bp,bburx=600bp,bbury=760bp,clip=}}
\caption{\label{dphqet}Measured values of 
$\left| {\cal F}(\tilde{w})V_{cb}\right|  $,  ${\cal F}(\tilde{w})$, where $\tilde{w}$ represents the
measured Lorentz invariant $w$. The 
results from the $\nu $ reconstruction analysis are shown as solid dots and 
the missing
mass analysis ones as asterisks. The combined results are shown as the solid 
line.
Note that the curve shown is not a fit to the data. The dashed line
denotes as measured in $B\dec D^{\star }\ell \nu$ 
decays.}
\end{figure}

The same data can be used to extract \vcb\ with a different method, 
inspired by HQET. In this approach the relevant dynamical variable is the 
velocity transfer $w$, related to $q^2$ by:
\begin{equation}
w=\frac{M_B^2+ M_D^2-q^2}{2M_B M_D}.
\end{equation}      
The point $w=1$ corresponds to the situation where the $B$ decays to a 
$D$ at rest in the $B$ frame.   Fig.~\ref{dphqet} shows  
the experimental
 data for $\left|{\cal F}(w)V_{cb}\right|$. In particular
it can be seen that it is difficult to ascertain the curvature of the Isgur-Wise universal 
function  because of the low statistics at the points close to $w=1$ and thus the 
uncertainty in the
extraction of \vcb\ must  reflect this additional uncertainty \cite{stone}. The
fit to these data gives ${\cal F}(1) = (3.46\pm 0.42\pm 0.46)\times 10^{-2}$. 

\begin{figure}[htbp]
\vspace{-0.8cm}
\centerline{\psfig{figure=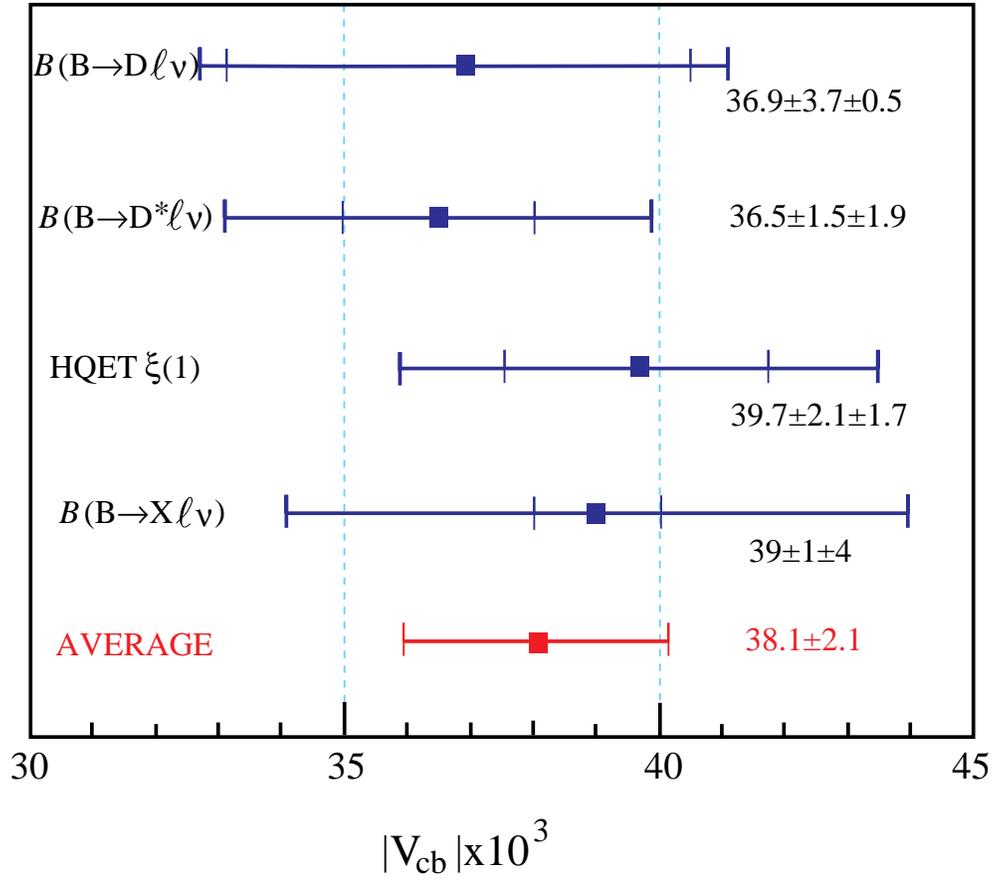,height=5.0in,bbllx=0bp,bblly=80bp,bburx=600bp,bbury=600bp,clip=}}
\vspace{-.2cm}
\caption[]{\label{Vcb} Results of four different methods used to evaluate 
$\left|V_{cb}\right|$, and the resulting average. The horizontal lines show the values,
the statistical errors out to the thin vertical lines, and the systematic errors
added on linearly out to the thick vertical lines.}
\end{figure}

In addition to the measurement discussed in this paper, the parameter \vcb\ has
been studied experimentally with several different methods. Many groups have
reported their findings for the branching fraction ${\cal B}(B\dec D^{\star} 
\ell
\bar{\nu})$\ \cite{dstarlnu} and related their measurement to several different
theoretical models to extract their estimate of \vcb . In addition this decay
can be studied with the HQET method discussed above \cite{hqet}.
This approach has a special interest from the point 
of view of HQET theorists because Luke's theorem \cite{luke} states that for
$w=1$, the mass dependent corrections vanish to first order and therefore 
the first non--zero term occurs with power $1/m_Q^2$. This 
implies that for a well behaved perturbative expansion, these corrections 
should play a minor role and thus allow an extraction of \vcb\ prone to 
smaller theoretical uncertainty. Lastly, the 
inclusive semileptonic decay can be related to \vcb\ \cite{stone}. The 
average values of \vcb\ obtained with these techniques are shown in 
Fig.~\ref{Vcb}.
The average value $\left| V_{cb}\right| =0.0381\pm 0.0021$, corresponding to 
the value for 
the 
$CKM$ parameter $A=0.784\pm 0.043$, has been obtained by adding statistical
 and systematic errors in the various estimates linearly and then adding the 
different 
methods in quadrature. This procedure is by no means rigorous but
it should give a conservative estimate of the final errors.  

\subsection{The CKM parameter \vub\ and charmless semileptonic decays}
The first evidence of a non-zero \vub\ was obtained by CLEO I \cite{bucleo}, by 
studying inclusive semileptonic decays. They reported an excess of leptons beyond
the kinematic endpoint for the decay $B\dec D \ell \bar{\nu}$. This result was quickly
confirmed by ARGUS \cite{argus} and then studied in more detail and with better
statistical  accuracy by CLEO II \cite{bucltwo} . There are two crucial
issues that make the extraction of \vub\ from experimental data trickier than the
extraction of \vcb . First of all, in this case we have the possibility of light hadronic systems recoiling
 against the lepton--$\bar{\nu}$ pair. Therefore the $q^2$ domain spanned by
these decays is much bigger and the assumed $q^2$ dependence of the form 
factors strongly affects the predicted rate $\gamma _u$ and fraction of high
momentum leptons $f_u(p)$. In addition, there is some uncertainty on the
composition of the hadronic system recoiling against the lepton-$\bar{\nu}$ pair.

\begin{figure}[htbp]
\vspace{-1.5cm}
\centerline{\psfig{figure=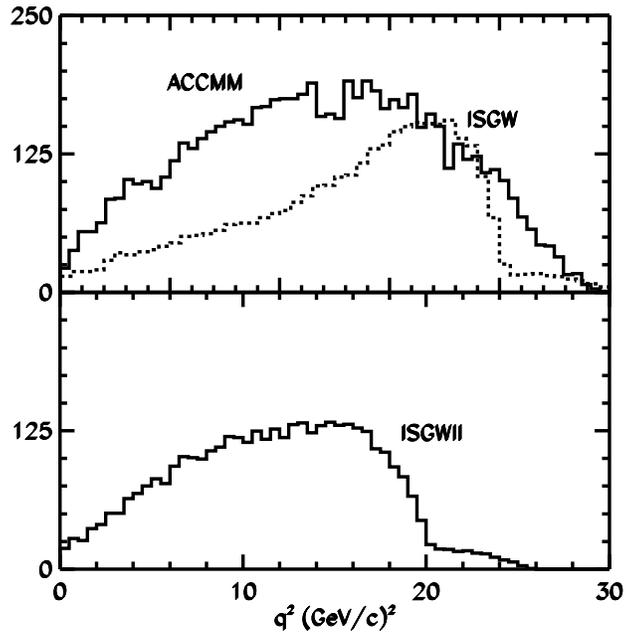,height=5.in,bbllx=0bp,bblly=0bp,bburx=600bp,bbury=700bp,clip=}}
\vspace{-2.2cm}
\caption[]{\label{qsq}$q^2$ distribution, for charmless semileptonic
$B$ decays in the model of Altarelli \etal (ACCMM) and the orginal ISGW model
shown on top, and the new ISGW II model shown on the bottom. The areas reflect
the predicted widths, but the vertical scale is arbitrary. The high $q^2$ tails
on the ISGW and ISGW II models arise from the $\pi\ell\nu$ final state.}
\end{figure}

Consequently models that focus on a few exclusive hadronic final states are not
likely to give reliable predictions for $\gamma _u$, as it is quite unlikely
that the whole Dalitz plot is dominated by the low lying resonances.   On 
the contrary, `inclusive' models, like the one
proposed by Altarelli et a. (ACCMM) \cite{acm},  based upon a quark-hadron 
duality picture,
become more relevant when several final states are involved.  The importance
of the theoretical uncertainty in the extraction of this parameter is 
illustrated by
the fact that the CLEO II estimate of \vubc\ changed by a factor of two 
depending
upon the model used \cite{bucltwo}. The theoretical uncertainty in the extraction of
this parameter is closely related to the  $q^2$  dependence of the form factors, as 
shown by Artuso \cite{artusoqsq} , by comparing the ACCMM  and ISGW \cite{isgw}
predictions for the $q^2$ distributions. Recently N. Isgur and D. Scora
have revised the ISGW model,  aiming at making it more realistic at high
\qsq\  and using some constraints on the form factors derived from HQET.  This
model, referred to as ISGW II \cite{isgwii}, is now much closer to ACCMM in 
several respects.
In particular,  Fig.~\ref{qsq} shows the the predicted \qsq\ distribution of 
events with
leptons in the momentum range of 2.4 to 2.6 GeV/c (the interval adopted in the
CLEO II analysis leading to the most precise value of \vub\ available so far). 
It can be seen that the \qsq\ distribution predicted by ISGW II is much softer 
than the one expected by ISGW.

\begin{figure}[htbp]
\vspace{-0.6cm}
\centerline{\psfig{figure=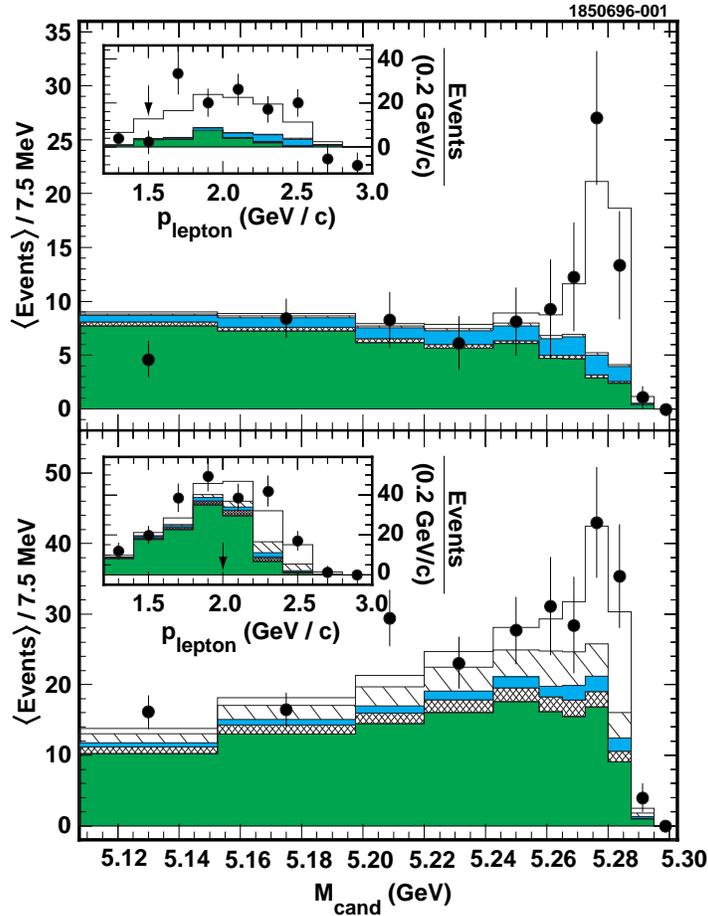,height=6.5in,bbllx=0bp,bblly=0bp,bburx=600bp,bbury=730bp,clip=}}
\vspace{-3.8cm}
\caption[]{\label{xlnu} The $B$ candidate mass distributions, $M_{cand}$,
for the sum of the scalar $\pi^+\ell\nu$ and $\pi^0\ell\nu$ (top) and the
vector modes ($\rho$ and $\omega$)$\ell \bar{\nu}$ (bottom). The points 
represent
the data after
continuum and fake background subtractions. The unshaded histogram is
the signal, while the dark shaded shows the $b\to c X$ background estimate,
the cross-hatched show the estimated $b\to u\ell\nu$ feed--down. For the 
$\pi$ (vector) modes, the light-shaded and hatched histograms are $\pi\to\pi$
(vector$\to$vector) and vector$\to\pi$ ($\pi\to$vector) cross--feeds,
respectively. The inserts show the lepton momentum spectra for the events
in the $B$ mass peak (the arrows indicate the momentum cuts).}
\end{figure}

It is obvious that in order to reduce the errors on the estimate of the parameter
\vub\  it is necessary to enlarge the set of experimental observables. In particular
the study of exclusive channels is the necessary first step to check in more detail
different theoretical predictions.  CLEO has recently measured the branching
fractions for exclusive charmless semileptonic decays involving a $\pi$, $\rho$
or $\omega$ meson in the final state \cite{cleorho}. Both charged and neutral $B$ decays have been
studied. The experimental technique adopted relies again upon the $\bar{\nu}$ 
energy
and momentum estimates from the rest of the event. The $\bar{\nu}$ invariant
mass squared is calculated from the reconstructed $E_{miss}$ and $\vec{p}_{miss}$
as $M_{miss}^2= E_{miss}^2 -\vec{p}_{miss}~^2$ and the selection criterion
$M^2_{miss}/2E_{miss} < 300 MeV$ is subsequently applied. This
approach allows to reconstruct these decays 
with the
techniques developed to study exclusive $B$ meson decays.   Fig.~\ref{xlnu}
 shows the
beam constrained invariant mass $M_{cand}$, evaluated as:
\begin{equation}
M^2_{cand}=E_{beam}^2 - (\vec{p}_{\bar{\nu }}+\vec{p}_l +\vec{p}_{(\pi \ {\rm or}\ 
\rho)} )^2.
\end{equation}

\begin{figure}[htbp]
\centerline{\psfig{figure=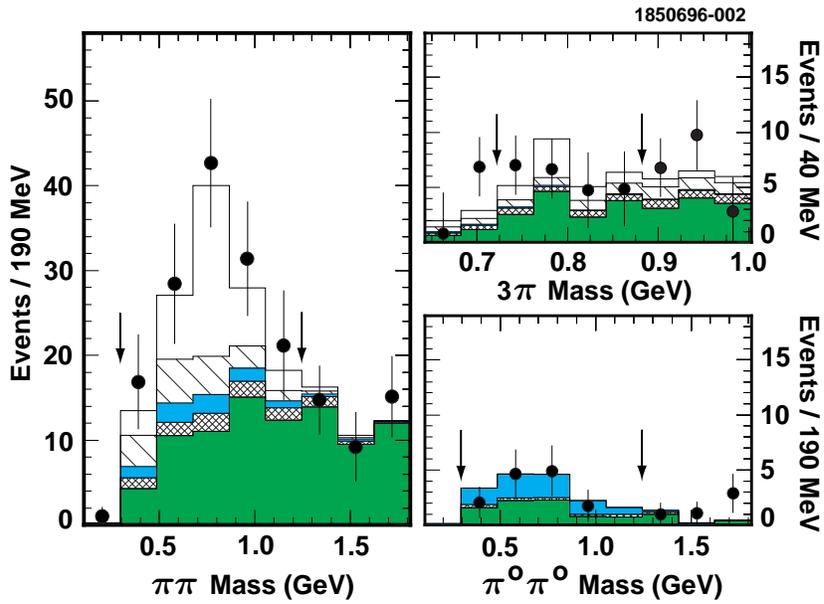,height=7.in,bbllx=0bp,bblly=0bp,bburx=600bp,bbury=700bp,clip=}}
\vspace{-9.1cm}
\caption[]{\label{rholnu} Mass distributions for $\pi^+\pi^-$ plus
$\pi^+\pi^0$ (left), $3\pi$ (upper right) and $\pi^0\pi^0$ (lower right),
for events which are candidates $B\to X\ell
\bar{\nu}$ decays which satisfy
all the other $B$ candidate cuts including a cut on the $B$ mass. The
shading is the same as on the previous figure. The arrows indicate the mass
range used in the analysis.}
\end{figure}

In the study of hadronic final states involving $\pi \pi$ in the final system it is often
difficult to prove that this system is indeed prevalently $\rho$. CLEO addresses this
question by plotting the $\pi^+\pi^0$ summed mass spectrum, as shown in 
 Fig.~\ref{rholnu}.
They also show the $\pi^0 \pi ^0$ case, which cannot be $\rho$. In the latter case
they see a flat background, while the former distribution show some peaking at the
expected nominal resonance mass. On the other hand the $3\pi$ spectrum show
little evidence of resonant $\omega$. There is clearly the need for more statistics,
but this analysis is performed under the  assumption that the resonant component
is dominant.

The five modes $B^-\dec \pi ^0  \ell \bar{\nu}$,  $B^-\dec \rho ^0 \ell 
\bar{\nu}$, 
$B^-\dec \omega ^0 \ell \bar{\nu}$, 
 $\bar{B}^0\dec \pi ^+ \ell \bar{\nu}$,  $\bar{B}^0\dec \rho ^+ \ell \bar{\nu}$ are fitted
simultaneously, using the constraints following from isospin:
\begin{eqnarray}
\Gamma(\bar{B}^0 \dec \pi^+ \ell \bar{\nu}) = 
\Gamma(B^- \dec \pi ^0  \ell \bar{\nu}), \\
\Gamma(\bar{B}^0 \dec \rho^+ \ell \bar{\nu}) = 
2\Gamma(B^- \dec \rho ^0  \ell \bar{\nu})=2
\Gamma(B^- \dec \omega ^0  \ell \bar{\nu}).
\end{eqnarray}
The analysis yields the branching fractions
${\cal B} (\bar{B}^0 \dec \pi ^+ \ell \bar{\nu}) = 
(1.8\pm 0.4 \pm 0.3\pm 0.2)\times 10^{-4}$ 
and
${\cal B} (\bar{B}^0 \dec \rho ^+ \ell \bar{\nu})= 
(2.5\pm 0.4 ^{+0.5}_{-0.7}\pm 0.5)\times 10^{-4}$. The ratio
between the partial widths to vector and pseudoscalar final state is 
interesting
because it provides a useful consistency check of the soundness of the 
assumptions adopted by different phenomenological models. Table 2 shows a
comparison between the theoretical ratio $\Gamma (B\dec \rho \ell
\bar{\nu})/\Gamma (B\dec \pi \ell \bar{\nu})$ predicted by a variety of  
quark model
calculations \cite{isgwii}, \cite{wbs}, \cite{ks}, \cite{meli}  and the corresponding measured 
values, using the same model 
to estimate the efficiency corrections. It can be seen that for some models 
there 
is a
quite significant discrepancy between the predicted and measured value. In
particular  the Korner and Schuler (KS) model has only a 0.5\% likelihood to 
be consistent with the data.

\begin{figure}[htbp]
\centerline{\psfig{figure=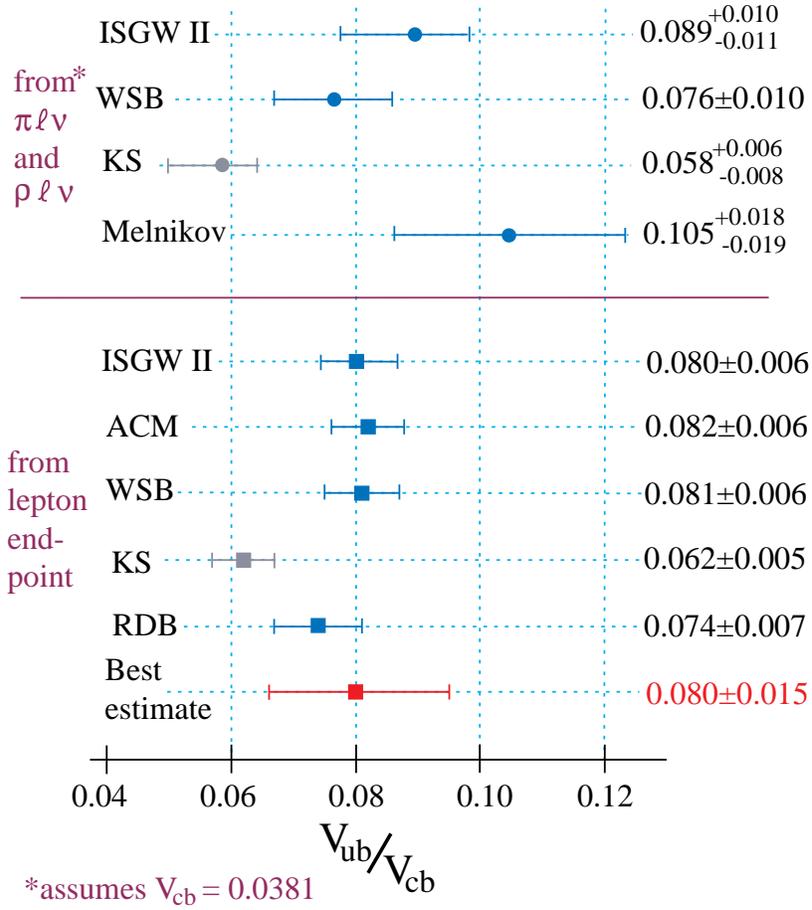,height=7.0in,bbllx=0bp,bblly=0bp,bburx=600bp,bbury=750bp,clip=}}
\vspace{-5.2cm}
\caption[]{\label{Vub_cb} Values of $\left| V_{ub}/V_{cb}\right|$ obtained from the
exclusive $\pi\ell\nu$ and $\rho\ell\nu$ analyses combined, taking
\vcb\ = 0.0381, and results from the inclusive endpoint analysis. The
best estimate combining all models except KS is also given.}
\end{figure}

 Fig.~\ref{Vub_cb} summarizes our present knowledge of the parameter $\left| V_{ub}/V_{cb}\right| $, both
from the exclusive and inclusive analyses. For the inclusive analysis results 
from CLEO I \cite{bucleo} and ARGUS \cite{argus} have been included in the 
average. 
The KS 
model has been excluded from the average as their 
prediction of the pseudoscalar to vector ratio is inconsistent with
the data. In addition, the ISGW model has been replaced by the updated ISGW II
model. The spread in the  $\left| V_{ub}/V_{cb}\right|$ estimates related to 
model dependence
is now narrowed compared to previous analyses. It should be noted that the 
models adopted are now much more similar in their estimate of the $q^2$ 
dependence of the form factors used and an experimental confirmation of the
correctness of this predicted dependence is eagerly awaited.
Furthermore, new approaches, based either on QCD sum rule techniques
\cite{ural}, \cite{ball} or on lattice QCD \cite{ukqcd} provide new 
theoretical perspectives
on these decay.
  Nonetheless at the present time  
the model dependence still dominates the errors. A conservative estimate 
gives:
\begin{equation}
\left| \frac{V_{ub}}{V_{cb}}\right| = 0.080 \pm 0.015,
\end{equation}
which corresponds to the constraint:
\begin{equation}
\rho ^2 + \eta ^2 = (0.36\pm 0.07)^2.
\end{equation}

\begin{figure}[h]
\vspace{-.2cm}
\centerline{\psfig{figure=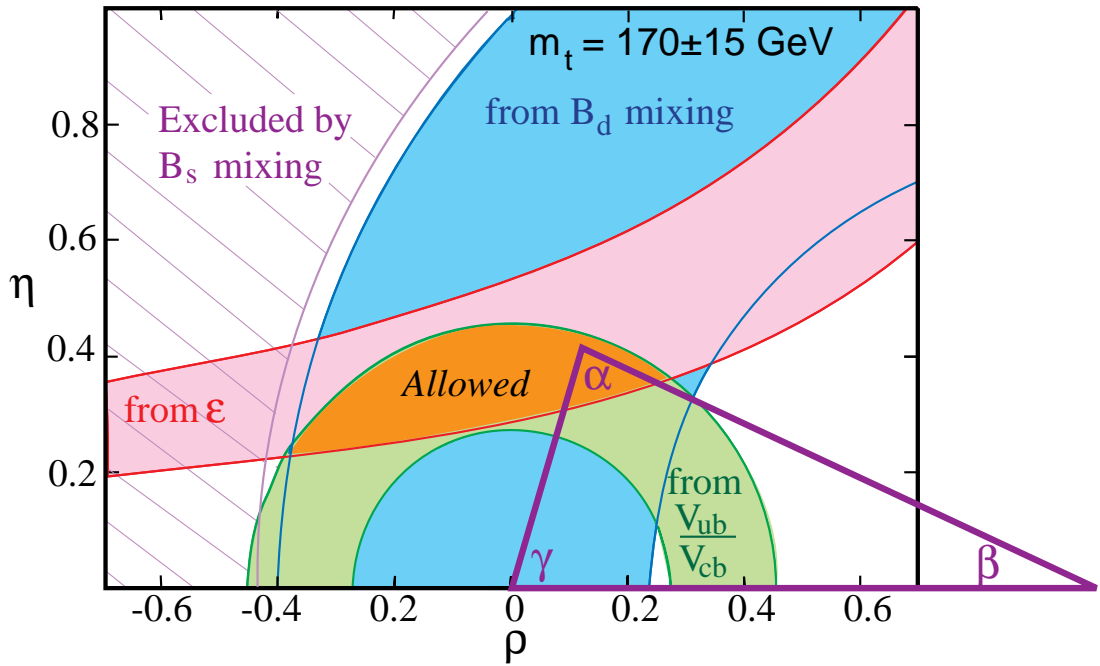,height=9.0in,bbllx=0bp,bblly=0bp,bburx=600bp,bbury=800bp,clip=}}
\vspace{-6.cm}
\caption[]{\label{ckm_fig} The regions in $\rho-\eta$ space (shaded) consistent
with measurements of CP violation in $K_L^0$ decay ($\epsilon$),
 $\left| V_{ub}/V_{cb}\right|$
in semileptonic $B$ decay, $B_d^0$ mixing, and the excluded region from
limits on $B_S^0$ mixing. The allowed region is defined by the overlap of
the 3 permitted areas, and is where the apex of the  CKM triangle  sits.}
\end{figure}

Fig.~\ref{ckm_fig}  shows the constraints to the unitarity triangle deriving from the 
values of \vcb\ and \vub\ reported in this paper together with the constraints
coming from \bbar\ mixing and $CP$ violation in the \kkbar\ system 
($\epsilon$) \cite{slsvi}.
The width of the $\epsilon$ band is mostly affected by the uncertainties
in the Wolfenstein parameter $A$,  the top quark mass $m_t$, the charm quark mass 
$m_c$ and
the correction factor for the vacuum insertion approximation $f_K$. The width of the \bbar\ 
mixing band is dominated by the uncertainty on the $B$ meson decay constant
$f_B$ here taken to be in the range $240 MeV>f_B>160 MeV$.

\section{Factorization test with the decay $\bar{B}^0\dec D^+\ \ell\bar{\nu}$}
The role
of factorization in theoretical predictions on exclusive hadronic decays is
multifaceted and has indeed been debated at length. In the early theoretical
studies of hadronic heavy flavor decays \cite{bws}, factorization was used as
an {\it ansatz}, by assuming that in energetic
two body decays direct formation of hadrons by a quark current is a useful
approximation and that the two currents `factorize', i.e. the 
Hamiltonian can be expressed as a product of two currents, one which couples a meson 
in the final
state with the decaying one and the other which produces the second meson
out of the vacuum \cite{bws}, \cite{bari}.  For instance, 
$B\dec D\pi$
could be expressed as:
\begin{equation}
<D^+\pi ^- \mid {\cal O}\mid \bar{B}^0> =\frac{G_F}{\sqrt{2}}
V_{cb}
V_{ud}^{\star} 
c_1 <\pi ^-\mid \bar{u}_L
\gamma ^{\mu}d_L\mid 0>
 <D^+\mid \bar{b}_L\gamma_{\mu} c_L\mid B>
\end{equation}
There is no good reason why factorization should hold in
all two--body hadronic $B$ decays. However some arguments based on 
color transparency, 
originally
proposed by Bjorken \cite{bj} and later by Dugan and Grinstein \cite{dg}, 
make it plausible
that
decays involving one heavy and one light meson in the final state
factorize. This is because the time scale of interaction between two mesons
in the final state is too short to allow appreciable gluon exchange between 
them.
Dugan and Grinstein base their arguments on perturbative QCD. The soundness
of their approach has been questioned by Aglietti \cite{agli}.
Therefore precise experimental tests of factorization in hadronic $B$ decays is quite 
valuable to disentangle this very complex issue.

A factorization test proposed by Bjorken is based on the observation that if Eq. 13
 is valid,
then we can related the amplitude $<D^+\mid \bar{b}_L\gamma_{\mu} c_L\mid B>$ 
to the corresponding
matrix element in semileptonic $B$ decays. This implies:
\begin{equation}
\frac{{\cal B}(\bar{B}^0 \dec D^+h ^-)}{d{\cal B}/dq^2(\bar{B}^0 \dec D^+ \ell 
\bar{\nu})\mid _{q^2=m_{h^-}^2}}=6\pi ^2 c_1^2f_{h^-}^2 \left| V_{ud}\right| ^2
\end{equation}
here $c_1$ is a calculable short distance QCD correction and $f_{h^-}$ is the light meson 
decay constant.

We can use the measured differential decay width $d{\cal B}/dq^2$ reported in this paper
with the exclusive branching fractions for $\bar{B}^0 \dec D^+\pi^-$ and $\bar{B}^0 \dec D^+
\rho ^-$ \cite{cleolong} and use the known $f_{\pi}$ and $f_{\rho}$ to extract an
experimental estimate of the parameter $c_1$. The experimental data can be compared 
with QCD calculations to assess the applicability of factorization to these hadronic decays.
The results of this comparison are shown in Table 3. It can be seen that the agreement 
is quite satisfactory.

\section{Conclusions}
$B$ decays continue to provide a wealth of information on the Standard Model. In particular,
the 
detailed study of $B$ meson semileptonic decays has given new insights on 
the quark mixing parameters and on properties of the strong interactions when heavy quarks
are involved in the decay. More detailed studies are forthcoming  
with the increased luminosity planned for the CLEO upgrade and, later, with the
\epm\ $B$ factories and dedicated $B$ experiments at hadron machines. Therefore
 most of
the uncertainties related to model dependence will be disentangled and the 
Standard Model
will have to withstand one of its more substantial challenges.

\section{Acknowledgements}
The author would like to acknowledge the contribution of CLEO and CESR scientists and technical staff to obtain the data presented in this paper. 
Many thanks are due to C. Sachrajda, G. Martinelli and N. Uraltsev for 
inspiring discussion during the Beauty '96 conference.  F. Ferroni deserves
gratitude for the impeccable organization  and the unforgettable visit to the Hadrian Villa and P. Schlein 
for his indefatigable work towards this
conference series. Lastly Sheldon Stone should be thanked for his insightful
comments and Julia Stone should be thanked for nice breaks
provided to my evening writing.

\newpage

\begin{table}[htb]
\centering
\begin{tabular}{|l|c|c|c|}
\hline
Model & $f_+(0)$ prediction & $|V_{cb}f_+(0)|\times 10^3$ & $|V_{cb}|\times 
10^3$\\\hline
WSB & 0.70 & $25.7\pm 1.4\pm 1.7$ & $37.3\pm 2.0 \pm 2.5$ \\
KS & 0.69 & $25.7\pm 1.4\pm 1.7$ & $36.7\pm 2.0 \pm 2.5$ \\  
Demchuk {\etal} & 0.68 & $24.8\pm 1.1\pm 1.6$ & $36.4\pm 1.6 \pm 2.4$ \\ 
\hline
\end{tabular}
\caption{Results of $\bar{B}^o\to D^+\ell^-\bar{\nu}$ analysis.}
\label{table:dpln}
\end{table}

\begin{table}[htb]
\centering
\begin{tabular}{|l|c|c|c|c|}
\hline
Model & ${\cal B}(B\to\pi\ell\nu)$ &${\cal B}(B\to\rho\ell\nu)$ & 
$\Gamma(\rho)/\Gamma(\pi)$ & $\Gamma(\rho)/\Gamma(\pi)$\\ 
&$\times 10^4$&$\times10^4$ &&predicted\\\hline 
ISGW II & $2.0\pm 0.5\pm 0.3$ &  $2.2\pm 0.4^{+0.4}_{-0.6}$ & 
$1.1^{+0.5+0.2}_{-0.3-0.3}$ & 1.47\\
WSB & $1.8\pm 0.5\pm 0.3$ & $2.8\pm 0.5^{+0.5}_{-0.8}$ &
$1.6^{+0.7+0.3}_{-0.5-0.4}$ & 3.51\\
KS &$1.9\pm 0.5\pm 0.3$ & $1.9\pm 0.3^{+0.4}_{-0.5}$  &
{\bf $1.0^{+0.5+0.2}_{-0.3-0.3}$} &  4.55\\
Melikhov$^{\dagger}$ &  $1.8\pm 0.4\pm 0.3\pm 0.2$ & 
$2.8\pm 0.5^{+0.5}_{-0.8}\pm 0.4$ & $1.6^{+0.7+0.3}_{-0.5-0.4}\pm 0.11$ 
& 1.53$\pm$0.15\\\hline
\multicolumn{5}{l}{$\dagger$ The 3rd error arises from uncertainties in 
  the estimated form-factors}   
\end{tabular}
\caption{ Results from exclusive semileptonic $b\to u$ transistions}
\label{table:vubex}
\end{table}

\begin{table}[htb]
\centering
\begin{tabular}{|l|c|c|c|c|c|}
\hline
$h^-$ & $f_h({\rm MeV})$& $d{\cal B}/dq^2$ & ${\cal B}(\bar{B}^0\dec D^+h^-)$ & $c_1^2({\rm exp})$ &
 $c_1^2({\rm th})$ \\
\hline
$\pi ^-$ & $131.7\pm 0.2$ & $0.35\pm 0.04\pm 0.04 $ & $0.29\pm 0.04\pm 0.03$ 
&  $0.85\pm 0.20$ & 1.06-1.32 \\
$\rho ^-$ & $215.0 \pm 4.0 $ & $0.33\pm 0.04\pm 0.04$ 
& $0.81\pm 0.11\pm 0.12$ & $0.94\pm 0.24$ & 1.06-1.32 \\
\hline
\end{tabular}
\caption{ Results of factorization tests.}
\label{table:fact}
\end{table}
\hfill
\clearpage
\end{document}